\documentclass[english]{article}
\usepackage[T1]{fontenc}
\usepackage[latin9]{inputenc}
\usepackage{geometry}
\usepackage{color}
\usepackage{float}
\usepackage{amsmath}
\usepackage{graphicx}

\makeatletter
\@ifundefined{date}{}{\date{}}
\usepackage{cite}

\makeatother

\usepackage{babel}
\begin{document}

\title{Linear optics based entanglement concentration protocols for Cluster-type
entangled coherent state}

\author{Mitali Sisodia$^{1}$\thanks{mitalisisodiyadc@gmail.com; }, Chitra
Shukla$^{2}$\thanks{shukla.chitra@mail.tsinghua.edu.cn; chitrashukla07@gmail.com},
Gui-Lu Long$^{2}$ \thanks{gllong@mail.tsinghua.edu.cn }}
\maketitle
\begin{center}
$^{1}$Jaypee Institute of Information Technology, A 10, Sector 62,
Noida, UP 201307, India and 
\par\end{center}

\begin{center}
$^{2}$State Key Laboratory of Low-Dimensional Quantum Physics and
Department of Physics, Tsinghua University, Beijing 100084, China
\par\end{center}
\begin{abstract}
We proposed two linear optics based entanglement concentration protocols
(ECPs) to obtain maximally entangled 4-mode Cluster-type entangled
coherent state (ECS) from less (partially) entangled Cluster-type
ECS. The first ECP is designed using a superposition of single-mode
coherent state with two unknown parameters, whereas the second ECP
is obtained using a superposition of single-mode coherent state and
a superposition of two-mode coherent state with four unknown parameters.
The success probabilities have been calculated for both the ECPs.
Necessary quantum circuits enabling future experimental realizations
of the proposed ECPs are provided using linear optical elements. Further,
the benefit of the proposed schemes is established in the context
of long distance quantum communication where photon loss is an obstruction.
\end{abstract}

\section{Introduction\label{sec:Introduction}}

Quantum entanglement has already been established as \textcolor{black}{a}n
important resource that can accomplish various quantum information
processing tasks. Almost all the existing applications of entangled
states, such as quantum teleportation \cite{tele}, dense coding \cite{dense_coding},
quantum key distribution \cite{bb84,ekert,b92,vaidman-goldenberg},
quantum key agreement \cite{QKA1,QKA2}, quantum secret sharing \cite{QSS},
quantum secure direct communication \cite{QSDC1,QSDC2,review,ping-pong,our_DSQC,dsqc-1,Ortho_QSDC,QSDC-Review},
and many more \cite{QD,AQD,QC} are achieved successfully by applying
maximally entangled states (MES). However, in the practical scenarios,
noise affects an MES during storing, transmitting and processing steps.
Consequently, entanglement degradation happens unavoidably. This often
leads to reduced communication efficiency and insecure quantum communication
channel. In brief, noise is the biggest obstacle in maintaining a
shared entangled state which is essential for quantum communication,
and it (noise) often transforms an MES to a non-MES. Hence, it's extremely
important to design schemes for recovering MES from non-MESs. For
a pure state such a scheme of recovering MES is referred to as ECP,
whereas for a mixed state, such a scheme is called entanglement purification
protocol (EPP) \cite{EPP_Bennet,EPP_Pan,EPP_yamamoto,EPP_Gonta,EPP_Zwerger1,EPP_Zwerger2,EPP4}.
In this paper, we will restrict ourselves to the study of ECP for
a particularly important continuous variable (CV) MES- 4-mode Cluster-type
entangled coherent state (ECS). Before, we describe the specific reasons
for selecting this state, it may be apt to note that in this work,
we have proposed two ECPs, and ECPs as they are extremly important
for quantum information processing in the noisy environment. Specifically,
In 1996, Bennett et al. first introduced an ECP based on Schmidt decomposition
\cite{ECP_Bennet}. Since then, many ECPs have been proposed to achieve
MES for different quantum states \cite{ECP_Bose,ECP_Shi,ECP_Zhao,ECP_Sheng1,ECP_Sheng2,ECP_Sheng3,ECP_Deng,Dhara_Cluster_ECP1,W-ECS,Our_ECP1,Our_ECP2,ECP_Wang,ECP_Sheng_2017,ECP_Sheng_2019}. 

Multi-partite Cluster state plays a significant role in several quantum
communication \cite{HQIS,HJRSP,BCST} and computation \cite{one-way-computation1,one-way-computation2}
tasks. Specifically, it has a great importance and a unique characteristic
due to its robust entanglement in noisy channels as compared to 2-qubit
Bell and 3-qubit Greenberger-Horne-Zeilinger (GHZ) quantum states.
In other words, Cluster state shows high level persistency of connectedness
and is hard to be destroyed by \textcolor{black}{a }single-bit measurement,
i.e., less susceptible to decoherence. Unfortunately, the Cluster
states still interact with the noise just like other multi-partite
states and consequently becomes less entangled. To avoid such noise
effect, several ECPs have been reported for Cluster states in different
forms \cite{Lan_Cluster_ECP,Zhao_Cluster_ECP,Cluster_ECP2,Cluster_ECP5,Cluster_ECP6,Cluster_ECP7,Cluster_ECP8,Long_ECP_cluster,Song_Cluster_2017}.
However, we choose to propose two ECPs for 4-mode Cluster-type ECS.
The motivation behind proposing ECPs for 4-mode Cluster-type ECS would
become clear in the following text. Initially, the quantum communication
protocols were proposed using discrete variable (DV). Nowadays, interest
in design and development of quantum communication protocols, has
been shifted from DV to continuous variable (CV) quantum cryptography
\cite{CV_QKD}. Actually, there exists a kind of ECS, with the property
of entanglement being encoded in CV, it is extensively attracting
a lot of attention \cite{Barry1,Barry2} in perform various quantum
information processing tasks. Recently, a few of the quantum cryptographic
tasks have appeared using CV. For example, CV QKD \cite{CV_QKD,CVQKD1,CVQKD2}
analogue of DV QKD \cite{bb84}. Subsequently, an ECP has been reported
for 4-qubit Cluster-type hyperentangled states (HES), which is a hyperentangled
state between polarization states and coherent state of qubit \cite{HES_Cluster_ECP}.
In practical applications of ECS, the maximally ECSs are usually the
necessary resources. It is to be noted that Sheng et al., \cite{W-ECS}
have shed light on the work of Nguyen et al.'s \cite{Nguan_Ba_An}
study of an optimal QIP using W-type ECS that exhibits the existence
of a quantum information protocol called remote symmetric entanglement,
which could be done only using W-type ECS. Similarly, we expect that
the maximally entangled Cluster-type ECS would be highly demanding
as the necessary resources in certain practical applications of ECS.
Interestingly, CV Cluster state has been utilized for quantum computation
\cite{QC_CV_Cluster1,QC_CV_Cluster2}, in fact the experimental realization
has also been proposed for CV Cluster state \cite{Experimental_CV_Cluster1,Experimental_CV_Cluster}.
Hence, Cluster state is extremely important in both DV and CV quantum
information tasks. The 4-mode Cluster-type ECS, which we have used
to propose our ECP is a CV Cluster state, that has potential applications
in quantum information. It is, therefore essential to design an ECP
for Cluster-type ECS for the smooth operation of quantum cryptography
protocols. To the best of our knowledge, no ECP has been proposed
for Cluster-type ECS. Therefore, we have proposed two ECPs to convert
partially entangled 4-mode Cluster-type ECS into the maximally entangled
4-mode Cluster-type ECS. 

The paper is organized as follows: In Sec. \ref{sec:ECP-for-partially}
and Sec. \ref{sec:ECP-for-partially-1}, we have described our two
ECPs to obtain maximally entangled 4-mode Cluster-type ECS from less
(partially) entangled Cluster-type ECS. The first ECP is designed
using a single-mode coherent state with two unknown parameters, whereas
the second ECP is obtained using a single-mode coherent state and
a two-mode coherent state with four unknown parameters. Further, in
Sec. \ref{sec:Success-probability}, the success probability has been
calculated and plotted in Fig. \ref{fig:(Color-online)-(a)} (a) and
Fig. \ref{fig:(Color-online)-(a)} (b) for two ECPs, respectively.
Finally, the conclusion has been drawn in Sec. \ref{sec:Conclusion}.

\section{ECP for partially entangled 4-mode Cluster-type ECS assisted with
a superposition of single-mode coherent state\label{sec:ECP-for-partially}}

First, we propose an ECP for 4-mode Cluster-type ECS using a superposition
of single-mode coherent state having two terms (two unknown coefficients).
Specifically, at the end of this protocol, we should obtain a maximally
entangled CV 4-mode Cluster-type ECS expressed as 

\begin{equation}
\begin{array}{lcl}
|\psi\rangle_{abcd} & = & \frac{1}{2}\left[\left(|\alpha\rangle_{a}|\alpha\rangle_{b}|\alpha\rangle_{c}|\alpha\rangle_{d}+|-\alpha\rangle_{a}|-\alpha\rangle_{b}|\alpha\rangle_{c}|\alpha\rangle_{d}\right)\right.\\
 & + & \left.\left(|\alpha\rangle_{a}|\alpha\rangle_{b}|-\alpha\rangle_{c}|-\alpha\rangle_{d}-|-\alpha\rangle_{a}|-\alpha\rangle_{b}|-\alpha\rangle_{c}|-\alpha\rangle_{d}\right)\right].
\end{array}\label{eq:Maximally_entangled_Cluster}
\end{equation}

\begin{figure}[H]
\begin{centering}
\includegraphics[scale=0.5]{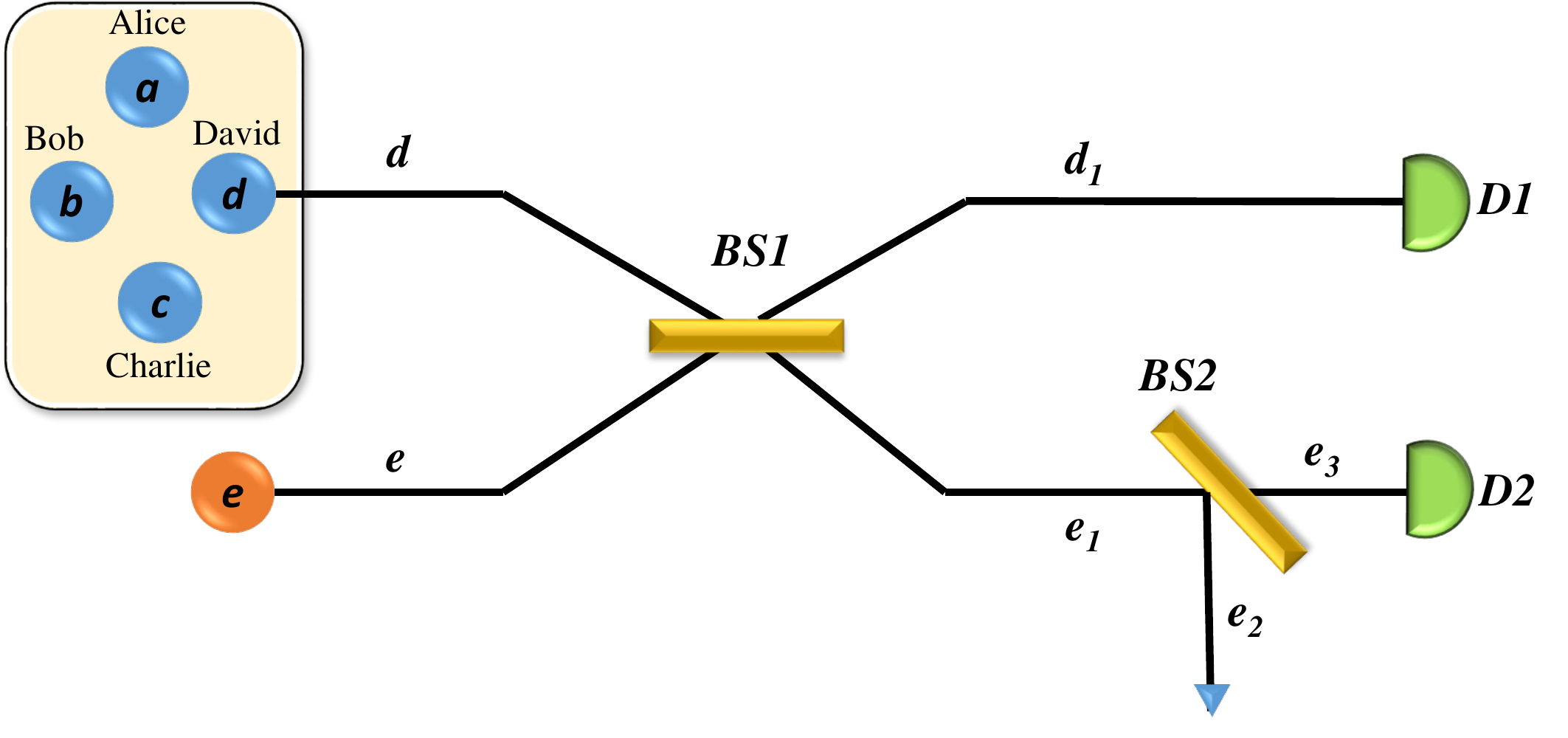}\caption{\label{fig:(Color-online)-The}(Color online) The schematic diagram
of the proposed 4-mode Cluster-type ECS has been shown where Alice,
Bob, Charlie and David initially share a partially entangled 4-mode
Cluster-type ECS. The subscripts $a,b,c$ and $d$ belong to Alice,
Bob, Charlie and David, respectively. Then, David prepares an ancilla
in the superposition of single-mode coherent state in spatial mode
$e$. The parity check measurement is performed using the 50:50 beam
splitter $BS1$. Further, to get the maximally entangled 4-mode Cluster-type
ECS, another $BS2$ is used which is helpful in reducing the increased
amplitude obtained in Eq. (\ref{eq:Postselcetion}).}
\par\end{centering}
\end{figure}

To perform entanglement concentration as shown in Fig. \ref{fig:(Color-online)-The},
we assume that Alice, Bob, Charlie and David used to share $|\psi\rangle_{abcd}$,
but due to noise this MES is transformed to a partially entangled
4-mode Cluster-type ECS of the form

\begin{equation}
\begin{array}{lcl}
|\Psi\rangle_{abcd} & = & N_{1}\left[\beta\left(|\alpha\rangle_{a}|\alpha\rangle_{b}|\alpha\rangle_{c}|\alpha\rangle_{d}+|-\alpha\rangle_{a}|-\alpha\rangle_{b}|\alpha\rangle_{c}|\alpha\rangle_{d}\right)\right.\\
 & + & \left.\gamma\left(|\alpha\rangle_{a}|\alpha\rangle_{b}|-\alpha\rangle_{c}|-\alpha\rangle_{d}-|-\alpha\rangle_{a}|-\alpha\rangle_{b}|-\alpha\rangle_{c}|-\alpha\rangle_{d}\right)\right],
\end{array}\label{eq:Partially_entangled_Cluster}
\end{equation}
where $N_{1}$ is the normalization coefficient, which can be written
as $N_{1}=\left[2\beta^{2}+2\gamma^{2}+2e^{-4|\alpha|^{2}}(\beta^{2}-\gamma^{2})\right]^{-\frac{1}{2}}$.
For simplicity, we assume $\beta$ and $\gamma$ are real numbers.
The subscripts $a,b,c$ and $d$ belong to Alice, Bob, Charlie and
David, respectively. Subsequently, David prepares an ancilla in the
superposition of single-mode coherent state in spatial mode $e$ of
the form

\begin{equation}
\begin{array}{ccc}
|\varPhi\rangle_{e} & = & N_{2}\left[\beta|\alpha\rangle_{e}+\gamma|-\alpha\rangle_{e}\right],\end{array}\label{eq:Single_coherent_state}
\end{equation}
where $N_{2}$ is the normalization coefficient expressed as $N_{2}=\left[\beta^{2}+\gamma^{2}+2\beta\gamma e^{-2|\alpha|^{2}}\right]^{-\frac{1}{2}}$.
Therefore, the combined state of the system can be expressed as 

\begin{equation}
\begin{array}{lcl}
|\xi\rangle_{abcde} & = & |\varPsi\rangle_{abcd}\otimes|\varPhi\rangle_{e}\\
 & = & N_{1}N_{2}\left[\beta\left(|\alpha\rangle_{a}|\alpha\rangle_{b}|\alpha\rangle_{c}|\alpha\rangle_{d}+|-\alpha\rangle_{a}|-\alpha\rangle_{b}|\alpha\rangle_{c}|\alpha\rangle_{d}\right)\right.\\
 & + & \left.\gamma\left(|\alpha\rangle_{a}|\alpha\rangle_{b}|-\alpha\rangle_{c}|-\alpha\rangle_{d}-|-\alpha\rangle_{a}|-\alpha\rangle_{b}|-\alpha\rangle_{c}|-\alpha\rangle_{d}\right)\right]\otimes\left[\beta|\alpha\rangle_{e}+\gamma|-\alpha\rangle_{e}\right],\\
|\xi\rangle_{abcde} & = & N_{1}N_{2}\left[\beta^{2}\left(|\alpha\rangle_{a}|\alpha\rangle_{b}|\alpha\rangle_{c}|\alpha\rangle_{d}|\alpha\rangle_{e}+|-\alpha\rangle_{a}|-\alpha\rangle_{b}|\alpha\rangle_{c}|\alpha\rangle_{d}|\alpha\rangle_{e}\right)\right.\\
 & + & \beta\gamma\left(|\alpha\rangle_{a}|\alpha\rangle_{b}|\alpha\rangle_{c}|\alpha\rangle_{d}|-\alpha\rangle_{e}+|-\alpha\rangle_{a}|-\alpha\rangle_{b}|\alpha\rangle_{c}|\alpha\rangle_{d}|-\alpha\rangle_{e}\right)\\
 & + & \gamma\beta\left(|\alpha\rangle_{a}|\alpha\rangle_{b}|-\alpha\rangle_{c}|-\alpha\rangle_{d}|\alpha\rangle_{e}-|-\alpha\rangle_{a}|-\alpha\rangle_{b}|-\alpha\rangle_{c}|-\alpha\rangle_{d}|\alpha\rangle_{e}\right)\\
 & + & \left.\gamma^{2}\left(|\alpha\rangle_{a}|\alpha\rangle_{b}|-\alpha\rangle_{c}|-\alpha\rangle_{d}|-\alpha\rangle_{e}-|-\alpha\rangle_{a}|-\alpha\rangle_{b}|-\alpha\rangle_{c}|-\alpha\rangle_{d}|-\alpha\rangle_{e}\right)\right].
\end{array}\label{eq:combined_state}
\end{equation}

\textcolor{red}{}

Further, they allow the spatial modes $d$ and $e$ to pass through
the 50:50 beam splitter $(BS)$ $(i.e.,$ $BS1)$. A $BS$ function
is to transform the two coherent states $|\alpha\rangle$ and $|\beta\rangle$
as shown below 

\begin{equation}
B_{12}|\alpha\rangle|\beta\rangle\rightarrow\frac{|\alpha+\beta\rangle}{\sqrt{2}}\frac{|\alpha-\beta\rangle}{\sqrt{2}}.\label{eq:BS_Function}
\end{equation}
When the photons in the spatial mode $d$ and $e$ incident on a $BS$,
there exists four different possibilities which can be expressed as 

\begin{equation}
\begin{array}{lcl}
|\alpha\rangle_{d}|\alpha\rangle_{e} & \rightarrow & |\sqrt{2}\alpha\rangle_{d_{1}}|0\rangle_{e_{1}},\\
|\alpha\rangle_{d}|-\alpha\rangle_{e} & \rightarrow & |0\rangle_{d_{1}}|\sqrt{2}\alpha\rangle_{e_{1}},\\
|-\alpha\rangle_{d}|\alpha\rangle_{e} & \rightarrow & |0\rangle_{d_{1}}|-\sqrt{2}\alpha\rangle_{e_{1}},\\
|-\alpha\rangle_{d}|-\alpha\rangle_{e} & \rightarrow & |-\sqrt{2}\alpha\rangle_{d_{1}}|0\rangle_{e_{1}}.
\end{array}\label{eq:Four_possibilities}
\end{equation}
Now, after the coherent states in the spatial modes $d$ and $e$
passing through the $BS1$, Eq.(\ref{eq:combined_state}) can be expressed
as 

\begin{equation}
\begin{array}{lcl}
|\xi\rangle_{abcd_{1}e_{1}} & = & N_{1}N_{2}\left[\beta^{2}\left(|\alpha\rangle_{a}|\alpha\rangle_{b}|\alpha\rangle_{c}|\sqrt{2}\alpha\rangle_{d_{1}}|0\rangle_{e_{1}}+|-\alpha\rangle_{a}|-\alpha\rangle_{b}|\alpha\rangle_{c}|\sqrt{2}\alpha\rangle_{d_{1}}|0\rangle_{e_{1}}\right)\right.\\
 & + & \beta\gamma\left(|\alpha\rangle_{a}|\alpha\rangle_{b}|\alpha\rangle_{c}|0\rangle_{d_{1}}|\sqrt{2}\alpha\rangle_{e_{1}}+|-\alpha\rangle_{a}|-\alpha\rangle_{b}|\alpha\rangle_{c}|0\rangle_{d_{1}}|\sqrt{2}\alpha\rangle_{e_{1}}\right)\\
 & + & \gamma\beta\left(|\alpha\rangle_{a}|\alpha\rangle_{b}|-\alpha\rangle_{c}|0\rangle_{d_{1}}|-\sqrt{2}\alpha\rangle_{e_{1}}-|-\alpha\rangle_{a}|-\alpha\rangle_{b}|-\alpha\rangle_{c}|0\rangle_{d_{1}}|-\sqrt{2}\alpha\rangle_{\boldsymbol{e_{1}}}\right)\\
 & + & \left.\gamma^{2}\left(|\alpha\rangle_{a}|\alpha\rangle_{b}|-\alpha\rangle_{c}|-\sqrt{2}\alpha\rangle_{d_{1}}|0\rangle_{e_{1}}-|-\alpha\rangle_{a}|-\alpha\rangle_{b}|-\alpha\rangle_{c}|-\sqrt{2}\alpha\rangle_{d_{1}}|0\rangle_{e_{1}}\right)\right].
\end{array}\label{eq:after_BS1}
\end{equation}
It is clear from Eq. (\ref{eq:after_BS1}) that components $|\alpha\rangle_{a}|\alpha\rangle_{b}|\alpha\rangle_{c}|\sqrt{2}\alpha\rangle_{d_{1}}|0\rangle_{e_{1}}$,
$|-\alpha\rangle_{a}|-\alpha\rangle_{b}|\alpha\rangle_{c}|\sqrt{2}\alpha\rangle_{d_{1}}|0\rangle_{e_{1}}$,
$|\alpha\rangle_{a}|\alpha\rangle_{b}|-\alpha\rangle_{c}|-\sqrt{2}\alpha\rangle_{d_{1}}|0\rangle_{e_{1}}$
and $-|-\alpha\rangle_{a}|-\alpha\rangle_{b}|-\alpha\rangle_{c}|-\sqrt{2}\alpha\rangle_{d_{1}}|0\rangle_{e_{1}}$
do not contain photon in the spatial mode $e_{1}$. However, the rest
of the four components do not contain photon in the spatial mode $d_{1}$.
Therefore, they can adopt the post selection method to select only
those four terms where the spatial mode $d_{1}$ has no photon which
can be written as 

\begin{equation}
\begin{array}{lcl}
|\xi\rangle_{abce_{1}} & = & N_{1}N_{2}\beta\gamma\left[|\alpha\rangle_{a}|\alpha\rangle_{b}|\alpha\rangle_{c}|\sqrt{2}\alpha\rangle_{e_{1}}+|-\alpha\rangle_{a}|-\alpha\rangle_{b}|\alpha\rangle_{c}|\sqrt{2}\alpha\rangle_{e_{1}}\right.\\
 & + & \left.|\alpha\rangle_{a}|\alpha\rangle_{b}|-\alpha\rangle_{c}|-\sqrt{2}\alpha\rangle_{e_{1}}-|-\alpha\rangle_{a}|-\alpha\rangle_{b}|-\alpha\rangle_{c}|-\sqrt{2}\alpha\rangle_{e_{1}}\right].
\end{array}\label{eq:Postselcetion}
\end{equation}

It is to be noted that the Eq. (\ref{eq:Postselcetion}) has the same
form with Eq. (\ref{eq:Maximally_entangled_Cluster}) with the only
difference that the amplitude of the coherent state in spatial mode
$e_{1}$ of Eq. (\ref{eq:Postselcetion}) is $\sqrt{2}$ times higher
than the coherent state in spatial mode $d$ of Eq. (\ref{eq:Maximally_entangled_Cluster}).
However, this $\sqrt{2}$ times increased amplitude in the above equation,
could be taken as advantage in long distance quantum communication,
where photon loss is an obstruction. Subsequently, in order to obtain
the maximally entangled 4-mode Cluster-type ECS as shown in Eq. (\ref{eq:Maximally_entangled_Cluster}),
the coherent state in spatial mode $e_{1}$ passes through the $BS2$
and we obtain

\begin{equation}
\begin{array}{lcl}
|\xi\rangle_{abce_{2}e_{3}} & = & N_{1}N_{2}\beta\gamma\left[|\alpha\rangle_{a}|\alpha\rangle_{b}|\alpha\rangle_{c}|\alpha\rangle_{e_{2}}|\alpha\rangle_{e_{3}}+|-\alpha\rangle_{a}|-\alpha\rangle_{b}|\alpha\rangle_{c}|\alpha\rangle_{e_{2}}|\alpha\rangle_{e_{3}}\right.\\
 & + & \left.|\alpha\rangle_{a}|\alpha\rangle_{b}|-\alpha\rangle_{c}|-\alpha\rangle_{e_{2}}|-\alpha\rangle_{e_{3}}-|-\alpha\rangle_{a}|-\alpha\rangle_{b}|-\alpha\rangle_{c}|-\alpha\rangle_{e_{2}}|-\alpha\rangle_{e_{3}}\right].
\end{array}\label{eq:After_BS2}
\end{equation}
Then, they perform the parity check measurement and can detect the
coherent state in the spatial mode $e_{2}$ (without making any distinction
between $|\alpha\rangle_{e_{2}}$ and $|-\alpha\rangle_{e_{2}}$)
and obtain the maximally entangled 4-mode Cluster-type ECS  of the
same form as shown in Eq. (\ref{eq:Maximally_entangled_Cluster}).

\begin{equation}
\begin{array}{lcl}
|\xi\rangle_{abce_{2}} & = & N\left[|\alpha\rangle_{a}|\alpha\rangle_{b}|\alpha\rangle_{c}|\alpha\rangle_{e_{2}}+|-\alpha\rangle_{a}|-\alpha\rangle_{b}|\alpha\rangle_{c}|\alpha\rangle_{e_{2}}\right.\\
 & + & \left.|\alpha\rangle_{a}|\alpha\rangle_{b}|-\alpha\rangle_{c}|-\alpha\rangle_{e_{2}}-|-\alpha\rangle_{a}|-\alpha\rangle_{b}|-\alpha\rangle_{c}|-\alpha\rangle_{e_{2}}\right],
\end{array}\label{eq:After_D2}
\end{equation}
where, $N=N_{1}N_{2}\beta\gamma$. Finally, they get the maximally
entangled 4-mode Cluster-type ECS as shown in Eq. (\ref{eq:Maximally_entangled_Cluster})
with success probability $P=4|N_{1}N_{2}\beta\gamma|^{^{2}}$ and
the same has been plotted in Fig. \ref{fig:(Color-online)-(a)} (a)
in Sec. \ref{sec:Success-probability}. 

\section{ECP for partially entangled 4-mode Cluster-type ECS assisted with
a superposition of single-mode coherent state and a superposition
of a two-mode coherent state\label{sec:ECP-for-partially-1}}

\begin{figure}[H]
\centering{}\includegraphics[scale=0.5]{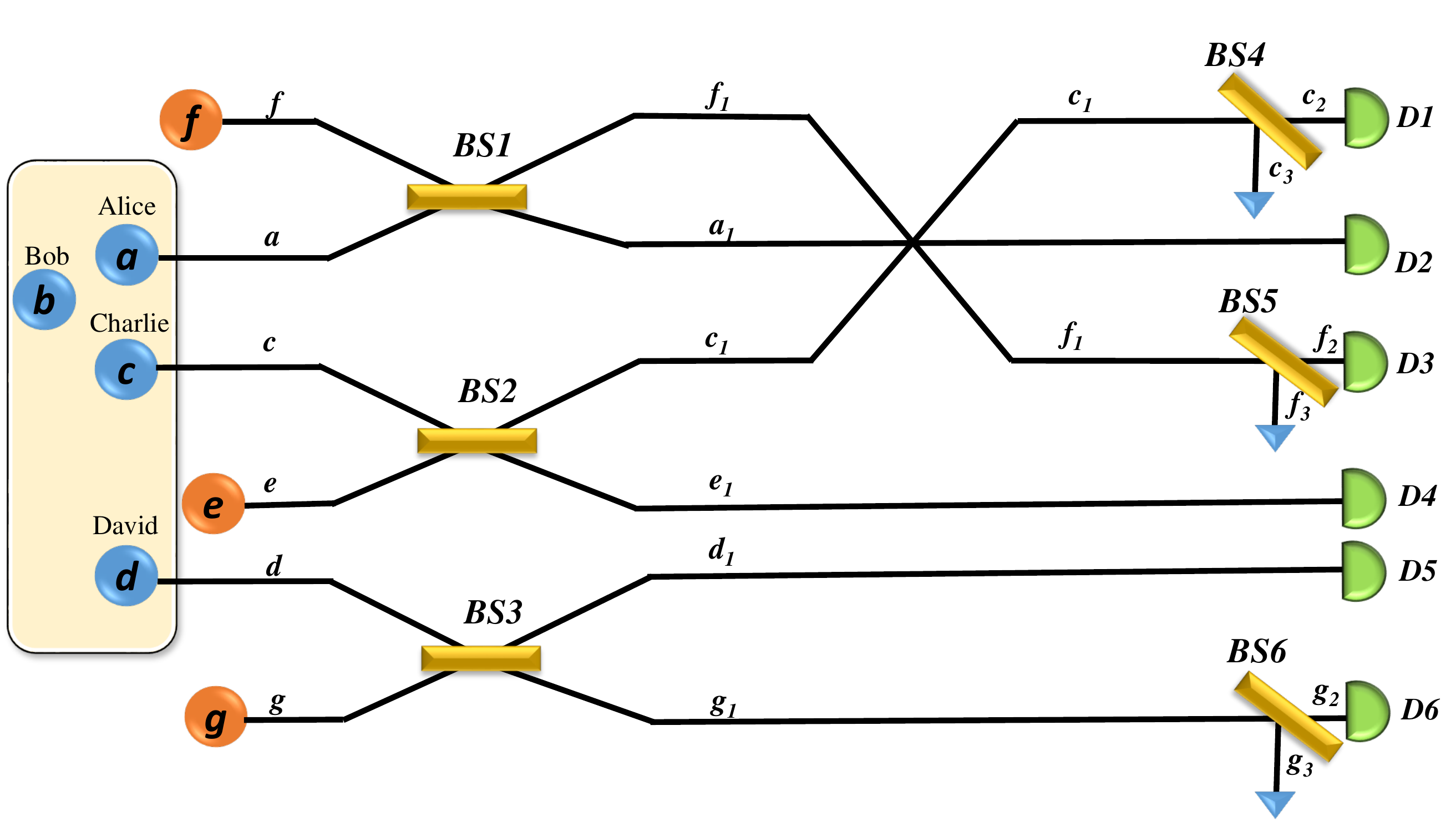}\caption{\label{fig:(Color-online)-The-1}(Color online) The schematic diagram
of the proposed 4-mode Cluster-type ECS has been shown where Alice,
Bob, Charlie and David initially share partially entangled 4-mode
Cluster-type ECS. The subscripts $a,b,c$ and $d$ belong to Alice,
Bob, Charlie and David, respectively. Then, David prepares an ancilla
in the superposition of single-mode coherent state in spatial mode
$g$ and a superposition of\textcolor{red}{{} }two-mode coherent state
in spatial mode $e$ and $f$. The parity check measurement is performed
using the 50:50 beam splitters $BS1,$ $BS2$ and $BS3$. Further,
to get the maximally entangled 4-mode Cluster-type ECS, $BS5$, $BS4$
and $BS6$, respectively, are used that is helpful in reducing the
increased amplitude obtained in Eq. (\ref{eq: after_swap}).}
\end{figure}

In this section, we propose an ECP for 4-mode Cluster-type ECS using
a superposition of single-mode coherent state and superposition of
a two-mode coherent state having four terms (four unknown coefficients).
As shown in Fig. \ref{fig:(Color-online)-The-1}, Alice, Bob, Charlie
and David share partially entangled 4-mode Cluster-type ECS of the
form

\begin{equation}
\begin{array}{lcl}
|\Psi^{\prime}\rangle_{abcd} & = & N_{3}\left[\beta|\alpha\rangle_{a}|\alpha\rangle_{b}|\alpha\rangle_{c}|\alpha\rangle_{d}+\gamma|-\alpha\rangle_{a}|-\alpha\rangle_{b}|\alpha\rangle_{c}|\alpha\rangle_{d}\right.\\
 & + & \left.\delta|\alpha\rangle_{a}|\alpha\rangle_{b}|-\alpha\rangle_{c}|-\alpha\rangle_{d}-\eta|-\alpha\rangle_{a}|-\alpha\rangle_{b}|-\alpha\rangle_{c}|-\alpha\rangle_{e_{2}}\right],
\end{array}\label{eq:arbitrary_cluster}
\end{equation}
where $N_{3}=\left[\beta^{2}+\gamma^{2}+\delta^{2}+\eta^{2}+2(\beta\gamma+\beta\delta-\gamma\eta-\delta\eta)e^{-4|\alpha|^{2}}+2(\delta\gamma-\eta\beta)e^{-8|\alpha|^{2}}\right]^{^{-\frac{1}{2}}}$
is the normalization coefficient. Here, we assume the coefficients
$\beta,\gamma,\delta$ and $\eta$ are real numbers. The subscripts
$a,b,c$ and $d$ belong to Alice, Bob, Charlie and David, respectively.
Further, David prepares the superposition of two-mode coherent state
in spatial mode $e$ and $f$ of the form 

\begin{equation}
\begin{array}{ccc}
|\Phi\rangle_{ef} & = & N_{4}\left[\beta|\alpha\rangle_{e}|\alpha\rangle_{f}+\gamma|\alpha\rangle_{e}|-\alpha\rangle_{f}+\delta|-\alpha\rangle_{e}|\alpha\rangle_{f}+\eta|-\alpha\rangle_{e}|-\alpha\rangle_{f}\right],\end{array}\label{eq: Two_qubit_ancillary}
\end{equation}
where $N_{4}=\left[\beta^{2}+\gamma^{2}+\delta^{2}+\eta^{2}+2(\beta\gamma+\beta\delta+\eta\gamma+\delta\eta)e^{-2|\alpha|^{2}}+2(\delta\gamma+\eta\beta)e^{-4|\alpha|^{2}}\right]^{-\frac{1}{2}}$
is the normalization coefficient. Therefore, the combined state of
the system can be expressed as

\begin{equation}
|\xi\rangle_{abcdef}=|\Psi^{\prime}\rangle_{abcd}\otimes|\Phi\rangle_{ef}\label{eq: Combined_state}
\end{equation}

\[
\begin{array}{lcl}
|\xi\rangle_{abcdef} & = & N_{3}N_{4}\left[\left[\beta|\alpha\rangle_{a}|\alpha\rangle_{b}|\alpha\rangle_{c}|\alpha\rangle_{d}+\gamma|-\alpha\rangle_{a}|-\alpha\rangle_{b}|\alpha\rangle_{c}|\alpha\rangle_{d}+\delta|\alpha\rangle_{a}|\alpha\rangle_{b}|-\alpha\rangle_{c}|-\alpha\rangle_{d}\right.\right.\\
 & - & \left.\eta|-\alpha\rangle_{a}|-\alpha\rangle_{b}|-\alpha\rangle_{c}|-\alpha\rangle_{e_{2}}\right]\otimes\left.\left[\beta|\alpha\rangle_{e}|\alpha\rangle_{f}+\gamma|\alpha\rangle_{e}|-\alpha\rangle_{f}+\delta|-\alpha\rangle_{e}|\alpha\rangle_{f}+\eta|-\alpha\rangle_{e}|-\alpha\rangle_{f}\right]\right],
\end{array}
\]

\[
\begin{array}{lcl}
|\xi\rangle_{abcdef} & = & N_{3}N_{4}\left[\beta^{2}|\alpha\rangle_{a}|\alpha\rangle_{b}|\alpha\rangle_{c}|\alpha\rangle_{d}|\alpha\rangle_{e}|\alpha\rangle_{f}+\gamma^{2}|-\alpha\rangle_{a}|-\alpha\rangle_{b}|\alpha\rangle_{c}|\alpha\rangle_{d}|\alpha\rangle_{e}|-\alpha\rangle_{f}\right.\\
 & + & \delta^{2}|\alpha\rangle_{a}|\alpha\rangle_{b}|-\alpha\rangle_{c}|-\alpha\rangle_{d}|-\alpha\rangle_{e}|\alpha\rangle_{f}-\eta^{2}|-\alpha\rangle_{a}|-\alpha\rangle_{b}|-\alpha\rangle_{c}|-\alpha\rangle_{e_{2}}|-\alpha\rangle_{e}|-\alpha\rangle_{f}\\
 & + & \gamma\eta(|-\alpha\rangle_{a}|-\alpha\rangle_{b}|\alpha\rangle_{c}|\alpha\rangle_{d}|-\alpha\rangle_{e}|-\alpha\rangle_{f}-|-\alpha\rangle_{a}|-\alpha\rangle_{b}|-\alpha\rangle_{c}|-\alpha\rangle_{e_{2}}|\alpha\rangle_{e}|-\alpha\rangle_{f})\\
 & + & \gamma\delta(|\alpha\rangle_{a}|\alpha\rangle_{b}|-\alpha\rangle_{c}|-\alpha\rangle_{d}|\alpha\rangle_{e}|-\alpha\rangle_{f}+|-\alpha\rangle_{a}|-\alpha\rangle_{b}|\alpha\rangle_{c}|\alpha\rangle_{d}|-\alpha\rangle_{e}|\alpha\rangle_{f})\\
 & + & \beta\gamma(|\alpha\rangle_{a}|\alpha\rangle_{b}|\alpha\rangle_{c}|\alpha\rangle_{d}|\alpha\rangle_{e}|-\alpha\rangle_{f}+|-\alpha\rangle_{a}|-\alpha\rangle_{b}|\alpha\rangle_{c}|\alpha\rangle_{d}|\alpha\rangle_{e}|\alpha\rangle_{f})\\
 & + & \delta\eta(|\alpha\rangle_{a}|\alpha\rangle_{b}|-\alpha\rangle_{c}|-\alpha\rangle_{d}|-\alpha\rangle_{e}|-\alpha\rangle_{f}-|-\alpha\rangle_{a}|-\alpha\rangle_{b}|-\alpha\rangle_{c}|-\alpha\rangle_{e_{2}}|-\alpha\rangle_{e}|\alpha\rangle_{f})\\
 & + & \beta\delta(|\alpha\rangle_{a}|\alpha\rangle_{b}|-\alpha\rangle_{c}|-\alpha\rangle_{d}|\alpha\rangle_{e}|\alpha\rangle_{f}+|\alpha\rangle_{a}|\alpha\rangle_{b}|\alpha\rangle_{c}|\alpha\rangle_{d}|-\alpha\rangle_{e}|\alpha\rangle_{f})\\
 & + & \left.\beta\eta(|\alpha\rangle_{a}|\alpha\rangle_{b}|\alpha\rangle_{c}|\alpha\rangle_{d}|-\alpha\rangle_{e}|-\alpha\rangle_{f}-|-\alpha\rangle_{a}|-\alpha\rangle_{b}|-\alpha\rangle_{c}|-\alpha\rangle_{e_{2}}|\alpha\rangle_{e}|\alpha\rangle_{f})\right].
\end{array}
\]
Subsequently, David also prepares\textcolor{red}{{} }one more ancilla
in the superposition of single-mode coherent state in spatial mode
$g$ of the form 

\begin{equation}
\begin{array}{ccc}
|\varPhi\rangle_{g} & = & N_{5}\left[\beta\gamma|\alpha\rangle_{g}+\delta\eta|-\alpha\rangle_{g}\right],\end{array}\label{eq: one_more_ancillary}
\end{equation}
where $N_{5}$ is the normalization coefficient and can be written
as $N_{5}=\left[\beta^{2}\gamma^{2}+\delta^{2}\eta^{2}+2\beta\gamma\delta\eta e^{-2|\alpha|^{2}}\right]^{-\frac{1}{2}}$.
The whole state of the system can be expressed as 

\[
|\zeta\rangle_{abcdefg}=|\xi\rangle_{abcdef}\otimes|\varPhi\rangle_{g},
\]
and same can be expanded as 

\begin{equation}
\begin{array}{lcl}
 &  & |\zeta\rangle_{abcdefg}\\
 & = & N_{3}N_{4}N_{5}\left[\beta^{3}\gamma|\alpha\rangle_{a}|\alpha\rangle_{b}|\alpha\rangle_{c}|\alpha\rangle_{d}|\alpha\rangle_{e}|\alpha\rangle_{f}|\alpha\rangle_{g}+\beta^{2}\delta\eta|\alpha\rangle_{a}|\alpha\rangle_{b}|\alpha\rangle_{c}|\alpha\rangle_{d}|\alpha\rangle_{e}|\alpha\rangle_{f}|-\alpha\rangle_{g}\right.\\
 & + & \beta\gamma^{3}|-\alpha\rangle_{a}|-\alpha\rangle_{b}|\alpha\rangle_{c}|\alpha\rangle_{d}|\alpha\rangle_{e}|-\alpha\rangle_{f}|\alpha\rangle_{g}+\gamma^{2}\delta\eta|-\alpha\rangle_{a}|-\alpha\rangle_{b}|\alpha\rangle_{c}|\alpha\rangle_{d}|\alpha\rangle_{e}|-\alpha\rangle_{f}|-\alpha\rangle_{g}\\
 & + & \delta^{2}\beta\gamma|\alpha\rangle_{a}|\alpha\rangle_{b}|-\alpha\rangle_{c}|-\alpha\rangle_{d}|-\alpha\rangle_{e}|\alpha\rangle_{f}|\alpha\rangle_{g}+\delta^{3}\eta|\alpha\rangle_{a}|\alpha\rangle_{b}|-\alpha\rangle_{c}|-\alpha\rangle_{d}|-\alpha\rangle_{e}|\alpha\rangle_{f}|-\alpha\rangle_{g}\\
 & - & \eta^{2}\beta\gamma|-\alpha\rangle_{a}|-\alpha\rangle_{b}|-\alpha\rangle_{c}|-\alpha\rangle_{e_{2}}|-\alpha\rangle_{e}|-\alpha\rangle_{f}|\alpha\rangle_{g}-\eta^{3}\delta|-\alpha\rangle_{a}|-\alpha\rangle_{b}|-\alpha\rangle_{c}|-\alpha\rangle_{e_{2}}|-\alpha\rangle_{e}|-\alpha\rangle_{f}|-\alpha\rangle_{g}\\
 & + & \beta\gamma^{2}\eta(|-\alpha\rangle_{a}|-\alpha\rangle_{b}|\alpha\rangle_{c}|\alpha\rangle_{d}|-\alpha\rangle_{e}|-\alpha\rangle_{f}|\alpha\rangle_{g}-|-\alpha\rangle_{a}|-\alpha\rangle_{b}|-\alpha\rangle_{c}|-\alpha\rangle_{e_{2}}|\alpha\rangle_{e}|-\alpha\rangle_{f}|\alpha\rangle_{g})\\
 & + & \delta\gamma\eta^{2}(|-\alpha\rangle_{a}|-\alpha\rangle_{b}|\alpha\rangle_{c}|\alpha\rangle_{d}|-\alpha\rangle_{e}|-\alpha\rangle_{f}|-\alpha\rangle_{g}-|-\alpha\rangle_{a}|-\alpha\rangle_{b}|-\alpha\rangle_{c}|-\alpha\rangle_{e_{2}}|\alpha\rangle_{e}|-\alpha\rangle_{f}|-\alpha\rangle_{g})\\
 & + & \beta\gamma^{2}\delta(|\alpha\rangle_{a}|\alpha\rangle_{b}|-\alpha\rangle_{c}|-\alpha\rangle_{d}|\alpha\rangle_{e}|-\alpha\rangle_{f}|\alpha\rangle_{g}+|-\alpha\rangle_{a}|-\alpha\rangle_{b}|\alpha\rangle_{c}|\alpha\rangle_{d}|-\alpha\rangle_{e}|\alpha\rangle_{f}|\alpha\rangle_{g})\\
 & + & \gamma\delta^{2}\eta(|\alpha\rangle_{a}|\alpha\rangle_{b}|-\alpha\rangle_{c}|-\alpha\rangle_{d}|\alpha\rangle_{e}|-\alpha\rangle_{f}|-\alpha\rangle_{g}+|-\alpha\rangle_{a}|-\alpha\rangle_{b}|\alpha\rangle_{c}|\alpha\rangle_{d}|-\alpha\rangle_{e}|\alpha\rangle_{f}|-\alpha\rangle_{g})\\
 & + & \beta^{2}\gamma^{2}(|\alpha\rangle_{a}|\alpha\rangle_{b}|\alpha\rangle_{c}|\alpha\rangle_{d}|\alpha\rangle_{e}|-\alpha\rangle_{f}|\alpha\rangle_{g}+|-\alpha\rangle_{a}|-\alpha\rangle_{b}|\alpha\rangle_{c}|\alpha\rangle_{d}|\alpha\rangle_{e}|\alpha\rangle_{f}|\alpha\rangle_{g})\\
 & + & \beta\gamma\delta\eta(|\alpha\rangle_{a}|\alpha\rangle_{b}|\alpha\rangle_{c}|\alpha\rangle_{d}|\alpha\rangle_{e}|-\alpha\rangle_{f}|-\alpha\rangle_{g}+|-\alpha\rangle_{a}|-\alpha\rangle_{b}|\alpha\rangle_{c}|\alpha\rangle_{d}|\alpha\rangle_{e}|\alpha\rangle_{f}|-\alpha\rangle_{g})\\
 & + & \beta\gamma\delta\eta(|\alpha\rangle_{a}|\alpha\rangle_{b}|-\alpha\rangle_{c}|-\alpha\rangle_{d}|-\alpha\rangle_{e}|-\alpha\rangle_{f}|\alpha\rangle_{g}-|-\alpha\rangle_{a}|-\alpha\rangle_{b}|-\alpha\rangle_{c}|-\alpha\rangle_{e_{2}}|-\alpha\rangle_{e}|\alpha\rangle_{f}|\alpha\rangle_{g})\\
 & + & \delta^{2}\eta^{2}(|\alpha\rangle_{a}|\alpha\rangle_{b}|-\alpha\rangle_{c}|-\alpha\rangle_{d}|-\alpha\rangle_{e}|-\alpha\rangle_{f}|-\alpha\rangle_{g}-|-\alpha\rangle_{a}|-\alpha\rangle_{b}|-\alpha\rangle_{c}|-\alpha\rangle_{e_{2}}|-\alpha\rangle_{e}|\alpha\rangle_{f}|-\alpha\rangle_{g})\\
 & + & \beta^{2}\gamma\delta(|\alpha\rangle_{a}|\alpha\rangle_{b}|-\alpha\rangle_{c}|-\alpha\rangle_{d}|\alpha\rangle_{e}|\alpha\rangle_{f}|\alpha\rangle_{g}+|\alpha\rangle_{a}|\alpha\rangle_{b}|\alpha\rangle_{c}|\alpha\rangle_{d}|-\alpha\rangle_{e}|\alpha\rangle_{f}|\alpha\rangle_{g})\\
 & + & \beta\delta^{2}\eta(|\alpha\rangle_{a}|\alpha\rangle_{b}|-\alpha\rangle_{c}|-\alpha\rangle_{d}|\alpha\rangle_{e}|\alpha\rangle_{f}|-\alpha\rangle_{g}+|\alpha\rangle_{a}|\alpha\rangle_{b}|\alpha\rangle_{c}|\alpha\rangle_{d}|-\alpha\rangle_{e}|\alpha\rangle_{f}|-\alpha\rangle_{g})\\
 & + & \beta^{2}\gamma\eta(|\alpha\rangle_{a}|\alpha\rangle_{b}|\alpha\rangle_{c}|\alpha\rangle_{d}|-\alpha\rangle_{e}|-\alpha\rangle_{f}|\alpha\rangle_{g}-|-\alpha\rangle_{a}|-\alpha\rangle_{b}|-\alpha\rangle_{c}|-\alpha\rangle_{e_{2}}|\alpha\rangle_{e}|\alpha\rangle_{f}|\alpha\rangle_{g})\\
 & + & \left.\beta\delta\eta^{2}(|\alpha\rangle_{a}|\alpha\rangle_{b}|\alpha\rangle_{c}|\alpha\rangle_{d}|-\alpha\rangle_{e}|-\alpha\rangle_{f}|-\alpha\rangle_{g}-|-\alpha\rangle_{a}|-\alpha\rangle_{b}|-\alpha\rangle_{c}|-\alpha\rangle_{e_{2}}|\alpha\rangle_{e}|\alpha\rangle_{f}|-\alpha\rangle_{g})\right].
\end{array}\label{eq: Whole_combined_state}
\end{equation}

They allow the photons in spatial mode $a$ and $f,$ $c$ and $e,$
$d$ and $g$ pass through the beam splitters $BS1,\,\,BS2,\,\,BS3$,
respectively. Then the state becomes 

\begin{equation}
\begin{array}[t]{l}
|\zeta\rangle_{a_{1}bc_{1}d_{1}e_{1}f_{1}g_{1}}\\
=N_{3}N_{4}N_{5}\left[\beta^{3}\gamma|\sqrt{2}\alpha\rangle_{a_{1}}|\alpha\rangle_{b}|\sqrt{2}\alpha\rangle_{c_{1}}|\sqrt{2}\alpha\rangle_{d_{1}}|0\rangle_{e_{1}}|0\rangle_{f_{1}}|0\rangle_{g_{1}}+\beta^{2}\delta\eta|\sqrt{2}\alpha\rangle_{a_{1}}|\alpha\rangle_{b}|\sqrt{2}\alpha\rangle_{c_{1}}|0\rangle_{d_{1}}|0\rangle_{e_{1}}|0\rangle_{f_{1}}|\sqrt{2}\alpha\rangle_{g_{1}}\right.\\
+\beta\gamma^{3}|-\sqrt{2}\alpha\rangle_{a_{1}}|-\alpha\rangle_{b}|\sqrt{2}\alpha\rangle_{c_{1}}|\sqrt{2}\alpha\rangle_{d_{1}}|0\rangle_{e_{1}}|0\rangle_{f_{1}}|0\rangle_{g_{1}}+\gamma^{2}\delta\eta|-\sqrt{2}\alpha\rangle_{a_{1}}|-\alpha\rangle_{b}|\sqrt{2}\alpha\rangle_{c}|0\rangle_{d}|0\rangle_{e}|0\rangle_{f_{1}}|\sqrt{2}\alpha\rangle_{g}\\
+\delta^{2}\beta\gamma|\sqrt{2}\alpha\rangle_{a_{1}}|\alpha\rangle_{b}|-\sqrt{2}\alpha\rangle_{c_{1}}|0\rangle_{d_{1}}|0\rangle_{e_{1}}|0\rangle_{f_{1}}|-\sqrt{2}\alpha\rangle_{g_{1}}+\delta^{3}\eta|\sqrt{2}\alpha\rangle_{a_{1}}|\alpha\rangle_{b}|-\sqrt{2}\alpha\rangle_{c_{1}}|-\sqrt{2}\alpha\rangle_{d_{1}}|0\rangle_{e_{1}}|0\rangle_{f}|0\rangle_{g}\\
-\eta^{2}\beta\gamma|-\sqrt{2}\alpha\rangle_{a_{1}}|-\alpha\rangle_{b}|-\sqrt{2}\alpha\rangle_{c_{1}}|0\rangle_{d_{1}}|0\rangle_{e_{1}}|0\rangle_{f}|-\sqrt{2}\alpha\rangle_{g}-\eta^{3}\delta|-\sqrt{2}\alpha\rangle_{a_{1}}|-\alpha\rangle_{b}|-\sqrt{2}\alpha\rangle_{c_{1}}|-\sqrt{2}\alpha\rangle_{d_{1}}|0\rangle_{e_{1}}|0\rangle_{f_{1}}|0\rangle_{g_{1}}\\
+\beta\gamma^{2}\eta(|-\sqrt{2}\alpha\rangle_{a_{1}}|-\alpha\rangle_{b}|0\rangle_{c_{1}}|\sqrt{2}\alpha\rangle_{d_{1}}|\sqrt{2}\alpha\rangle_{e_{1}}|0\rangle_{f_{1}}|0\rangle_{g_{1}}-|-\sqrt{2}\alpha\rangle_{a_{1}}|-\alpha\rangle_{b}|0\rangle_{c_{1}}|0\rangle_{d_{1}}|-\sqrt{2}\alpha\rangle_{e_{1}}|0\rangle_{f_{1}}|-\sqrt{2}\alpha\rangle_{g_{1}})\\
+\delta\gamma\eta^{2}(|-\sqrt{2}\alpha\rangle_{a_{1}}|-\alpha\rangle_{b}|0\rangle_{c_{1}}|0\rangle_{d_{1}}|\sqrt{2}\alpha\rangle_{e_{1}}|0\rangle_{f_{1}}|\sqrt{2}\alpha\rangle_{g_{1}}-|-\sqrt{2}\alpha\rangle_{a_{1}}|-\alpha\rangle_{b}|0\rangle_{c_{1}}|-\sqrt{2}\alpha\rangle_{d_{1}}|-\sqrt{2}\alpha\rangle_{e_{1}}|0\rangle_{f_{1}}|0\rangle_{g_{1}})\\
+\beta\gamma^{2}\delta(|0\rangle_{a_{1}}|\alpha\rangle_{b}|0\rangle_{c_{1}}|0\rangle_{d_{1}}|-\sqrt{2}\alpha\rangle_{e_{1}}|\sqrt{2}\alpha\rangle_{f_{1}}|-\sqrt{2}\alpha\rangle_{g_{1}}+|0\rangle_{a_{1}}|-\alpha\rangle_{b}|0\rangle_{c_{1}}|\sqrt{2}\alpha\rangle_{d_{1}}|\sqrt{2}\alpha\rangle_{e_{1}}|-\sqrt{2}\alpha\rangle_{f_{1}}|0\rangle_{g_{1}})\\
+\gamma\delta^{2}\eta(|0\rangle_{a_{1}}|\alpha\rangle_{b}|0\rangle_{c_{1}}|-\sqrt{2}\alpha\rangle_{d_{1}}|-\sqrt{2}\alpha\rangle_{e_{1}}|\sqrt{2}\alpha\rangle_{f_{1}}|0\rangle_{g_{1}}+|0\rangle_{a_{1}}|-\alpha\rangle_{b}|0\rangle_{c_{1}}|0\rangle_{d_{1}}|\sqrt{2}\alpha\rangle_{e_{1}}|-\sqrt{2}\alpha\rangle_{f_{1}}|\sqrt{2}\alpha\rangle_{g_{1}})\\
+\beta^{2}\gamma^{2}(|0\rangle_{a_{1}}|\alpha\rangle_{b}|\sqrt{2}\alpha\rangle_{c_{1}}|\sqrt{2}\alpha\rangle_{d_{1}}|0\rangle_{e_{1}}|\sqrt{2}\alpha\rangle_{f_{1}}|0\rangle_{g_{1}}+|0\rangle_{a_{1}}|-\alpha\rangle_{b}|\sqrt{2}\alpha\rangle_{c_{1}}|\sqrt{2}\alpha\rangle_{d_{1}}|0\rangle_{e_{1}}|-\sqrt{2}\alpha\rangle_{f_{1}}|0\rangle_{g_{1}})\\
+\beta\gamma\delta\eta(|0\rangle_{a_{1}}|\alpha\rangle_{b}|\sqrt{2}\alpha\rangle_{c_{1}}|0\rangle_{d_{1}}|0\rangle_{e_{1}}|\sqrt{2}\alpha\rangle_{f_{1}}|\sqrt{2}\alpha\rangle_{g_{1}}+|0\rangle_{a_{1}}|-\alpha\rangle_{b}|\sqrt{2}\alpha\rangle_{c_{1}}|0\rangle_{d_{1}}|0\rangle_{e_{1}}|-\sqrt{2}\alpha\rangle_{f_{1}}|\sqrt{2}\alpha\rangle_{g_{1}})\\
+\beta\gamma\delta\eta(|0\rangle_{a_{1}}|\alpha\rangle_{b}|-\sqrt{2}\alpha\rangle_{c_{1}}|0\rangle_{d_{1}}|0\rangle_{e_{1}}|\sqrt{2}\alpha\rangle_{f_{1}}|-\sqrt{2}\alpha\rangle_{g_{1}}-|0\rangle_{a_{1}}|-\alpha\rangle_{b}|-\sqrt{2}\alpha\rangle_{c_{1}}|0\rangle_{d_{1}}|0\rangle_{e_{1}}|-\sqrt{2}\alpha\rangle_{f_{1}}|-\sqrt{2}\alpha\rangle_{g_{1}})\\
+\delta^{2}\eta^{2}(|0\rangle_{a_{1}}|\alpha\rangle_{b}|-\sqrt{2}\alpha\rangle_{c_{1}}|-\sqrt{2}\alpha\rangle_{d_{1}}|0\rangle_{e_{1}}|\sqrt{2}\alpha\rangle_{f_{1}}|0\rangle_{g_{1}}-|0\rangle_{a}|-\alpha\rangle_{b}|-\sqrt{2}\alpha\rangle_{c_{1}}|-\sqrt{2}\alpha\rangle_{d_{1}}|0\rangle_{e_{1}}|-\sqrt{2}\alpha\rangle_{f_{1}}|0\rangle_{g_{1}})\\
+\beta^{2}\gamma\delta(|\sqrt{2}\alpha\rangle_{a_{1}}|\alpha\rangle_{b}|0\rangle_{c_{1}}|0\rangle_{d_{1}}|-\sqrt{2}\alpha\rangle_{e_{1}}|0\rangle_{f_{1}}|-\sqrt{2}\rangle_{g_{1}}+|\sqrt{2}\alpha\rangle_{a_{1}}|\alpha\rangle_{b}|0\rangle_{c_{1}}|\sqrt{2}\alpha\rangle_{d_{1}}|\sqrt{2}\alpha\rangle_{e_{1}}|0\rangle_{f_{1}}|0\rangle_{g_{1}})\\
+\beta\delta^{2}\eta(|\sqrt{2}\alpha\rangle_{a_{1}}|\alpha\rangle_{b}|0\rangle_{c_{1}}|-\sqrt{2}\alpha\rangle_{d_{1}}|-\sqrt{2}\alpha\rangle_{e_{1}}|0\rangle_{f_{1}}|0\rangle_{g_{1}}+|\sqrt{2}\alpha\rangle_{a_{1}}|\alpha\rangle_{b}|0\rangle_{c_{1}}|0\rangle_{d_{1}}|\sqrt{2}\alpha\rangle_{e_{1}}|0\rangle_{f_{1}}|\sqrt{2}\alpha\rangle_{g_{1}})\\
+\beta^{2}\gamma\eta(|0\rangle_{a_{1}}|\alpha\rangle_{b}|0\rangle_{c_{1}}|\sqrt{2}\alpha\rangle_{d_{1}}|\sqrt{2}\alpha\rangle_{e_{1}}|\sqrt{2}\alpha\rangle_{f_{1}}|0\rangle_{g_{1}}-|0\rangle_{a_{1}}|-\alpha\rangle_{b}|0\rangle_{c_{1}}|0\rangle_{d_{1}`}|-\sqrt{2}\alpha\rangle_{e_{1}}|-\sqrt{2}\alpha\rangle_{f_{1}}|-\sqrt{2}\alpha\rangle_{g_{1}})\\
+\left.\beta\delta\eta^{2}(|0\rangle_{a_{1}}|\alpha\rangle_{b}|0\rangle_{c_{1}}|0\rangle_{d_{1}}|\sqrt{2}\alpha\rangle_{e_{1}}|\sqrt{2}\alpha\rangle_{f_{1}}|\sqrt{2}\alpha\rangle_{g_{1}}-|0\rangle_{a_{1}}|-\alpha\rangle_{b}|0\rangle_{c_{1}}|-\sqrt{2}\alpha\rangle_{d_{1}}|-\sqrt{2}\alpha\rangle_{e_{1}}|-\sqrt{2}\alpha\rangle_{f_{1}}|0\rangle_{g_{1}})\right].
\end{array}\label{eq: After_BS1_BS2_BS3}
\end{equation}
Further, they can adopt the post selection method to select the cases,
when the spatial mode $a_{1,\,}d_{1,\,}e_{1}$ have no photons, then
the state can be written as

\begin{equation}
\begin{array}{lcl}
|\zeta\rangle_{bc_{1}f_{1}g_{1}} & = & N_{3}N_{4}N_{5}\beta\gamma\delta\eta\left[|\alpha\rangle_{b}|\sqrt{2}\alpha\rangle_{c_{1}}|\sqrt{2}\alpha\rangle_{f_{1}}|\sqrt{2}\alpha\rangle_{g_{1}}+|-\alpha\rangle_{b}|\sqrt{2}\alpha\rangle_{c_{1}}|-\sqrt{2}\alpha\rangle_{f_{1}}|\sqrt{2}\alpha\rangle_{g_{1}}\right.\\
 & + & \left.|\alpha\rangle_{b}|-\sqrt{2}\alpha\rangle_{c_{1}}|\sqrt{2}\alpha\rangle_{f_{1}}|-\sqrt{2}\alpha\rangle_{g_{1}}-|-\alpha\rangle_{b}|-\sqrt{2}\alpha\rangle_{c_{1}}|-\sqrt{2}\alpha\rangle_{f_{1}}|-\sqrt{2}\alpha\rangle_{g_{1}}\right].
\end{array}\label{eq: Postselection}
\end{equation}
Afterwards, they swap the mode $c_{1}\longleftrightarrow f_{1}$ and
the state becomes

\begin{equation}
\begin{array}{ccc}
|\zeta\rangle_{bf_{1}c_{1}g_{1}} & = & N_{3}N_{4}N_{5}\beta\gamma\delta\eta\left[|\alpha\rangle_{b}|\sqrt{2}\alpha\rangle_{f_{1}}|\sqrt{2}\alpha\rangle_{c_{1}}|\sqrt{2}\alpha\rangle_{g_{1}}+|-\alpha\rangle_{b}|-\sqrt{2}\alpha\rangle_{f_{1}}|\sqrt{2}\alpha\rangle_{c_{1}}|\sqrt{2}\alpha\rangle_{g_{1}}\right.\\
 & + & \left.|\alpha\rangle_{b}|\sqrt{2}\alpha\rangle_{f_{1}}|-\sqrt{2}\alpha\rangle_{c_{1}}|-\sqrt{2}\alpha\rangle_{g_{1}}-|-\alpha\rangle_{b}|-\sqrt{2}\alpha\rangle_{f_{1}}|-\sqrt{2}\alpha\rangle_{c_{1}}|-\sqrt{2}\alpha\rangle_{g_{1}}\right].
\end{array}\label{eq: after_swap}
\end{equation}
Now, we can see that the Eq. (\ref{eq: after_swap}) has the same
form with Eq. (\ref{eq:Maximally_entangled_Cluster}) with the only
difference that the amplitude of the coherent states in spatial modes
$f_{1}$, $c_{1}$ and $g_{1}$ in Eq. (\ref{eq: after_swap}) are
$\sqrt{2}$ times higher than the coherent states in spatial modes
$b$, $c,$ and $d$, respectively in Eq. (\ref{eq:Maximally_entangled_Cluster}).
However, in order to obtain the maximally entangled 4-mode Cluster-type
ECS as shown in Eq. (\ref{eq:Maximally_entangled_Cluster}), the photons
in spatial modes $c_{1,\,}f_{1}$ and $g_{1}$ pass through the beam
splitters $BS4,\,\,BS5,\,\,BS6$ and then the state can be expressed
as

\begin{equation}
\begin{array}[t]{l}
|\zeta\rangle_{bf_{2}f_{3}c_{2}c_{3}g_{2}g_{3}}\\
=N_{3}N_{4}N_{5}\beta\gamma\delta\eta\left[|\alpha\rangle_{b}|\alpha\rangle_{f_{2}}|\alpha\rangle_{f_{3}}|\alpha\rangle_{c_{2}}|\alpha\rangle_{c_{3}}|\alpha\rangle_{g_{2}}|\alpha\rangle_{g_{3}}+|-\alpha\rangle_{b}|-\alpha\rangle_{f_{2}}|-\alpha\rangle_{f_{3}}|\alpha\rangle_{c_{2}}|\alpha\rangle_{c_{3}}|\alpha\rangle_{g_{2}}|\alpha\rangle_{g_{3}}\right.\\
+\left.|\alpha\rangle_{b}|\alpha\rangle_{f_{2}}|\alpha\rangle_{f_{3}}|-\alpha\rangle_{c_{2}}|-\alpha\rangle_{c_{3}}|-\alpha\rangle_{g_{2}}|-\alpha\rangle_{g_{3}}-|-\alpha\rangle_{b}|-\alpha\rangle_{f_{2}}|-\alpha\rangle_{f_{3}}|-\alpha\rangle_{c_{2}}|-\alpha\rangle_{c_{3}}|-\alpha\rangle_{g_{2}}|-\alpha\rangle_{g_{3}}\right].
\end{array}\label{eq: After_BS4_BS5_BS6}
\end{equation}
After performing the parity check measurement, they can detect the
coherent state in the spatial modes $f_{3,\,}c_{3,\,}$ and $g_{3}$,
respectively (without making any distinction between $|\alpha\rangle_{f_{3}}$
and $|-\alpha\rangle_{f_{3}}$, $|\alpha\rangle_{c_{3}}$ and $|-\alpha\rangle_{c_{3}}$,
$|\alpha\rangle_{g_{3}}$ and $|-\alpha\rangle_{g_{3}}$) and the
state becomes 

\begin{equation}
\begin{array}{lcl}
|\zeta\rangle_{bf_{3}c_{3}g_{3}} & = & N^{'}\left[|\alpha\rangle_{b}|\alpha\rangle_{f_{3}}|\alpha\rangle_{c_{3}}|\alpha\rangle_{g_{3}}+|-\alpha\rangle_{b}|-\alpha\rangle_{f_{3}}|\alpha\rangle_{c_{3}}|\alpha\rangle_{g_{3}}\right.\\
 & + & \left.|\alpha\rangle_{b}|\alpha\rangle_{f_{3}}|-\alpha\rangle_{c_{3}}|-\alpha\rangle_{g_{3}}-|-\alpha\rangle_{b}|-\alpha\rangle_{f_{3}}|-\alpha\rangle_{c_{3}}|-\alpha\rangle_{g_{3}}\right],
\end{array}\label{eq:After_detection}
\end{equation}
where, $N^{'}=N_{3}N_{4}N_{5}\beta\gamma\delta\eta.$ Finally, they
obtain the maximally entangled 4-mode Cluster-type ECS of the same
form as shown in Eq. (\ref{eq:Maximally_entangled_Cluster}) with
success probability, $P=4|N_{3}N_{4}N_{5}\beta\gamma\delta\eta|^{^{2}}$
and the same has been plotted in Fig. \ref{fig:(Color-online)-(a)}
(b) in Sec. \ref{sec:Success-probability}.

\section{Success probability\label{sec:Success-probability}}

\begin{figure}[H]
\begin{centering}
\includegraphics[scale=0.65]{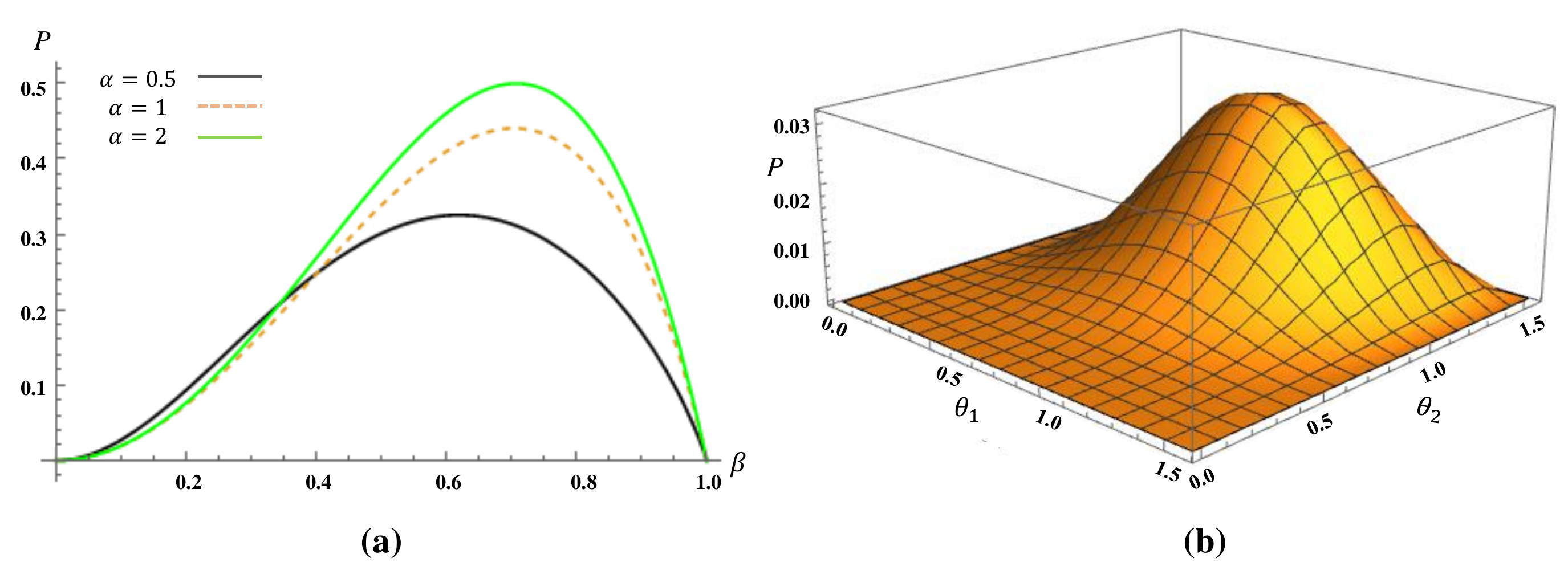}
\par\end{centering}
\centering{}\caption{\label{fig:(Color-online)-(a)}(Color online) (a) 2D Plot shows the
variation of success probability $P$ with $\beta$ for the first
ECP. Here, we choose $\alpha=0.5,$ $1,$ and $2$. (b) 3D Plot shows
the variation of success probability $P$ with $\theta_{1}\{0\rightarrow\frac{\pi}{2}\}$
and $\theta_{2}\{0\rightarrow\frac{\pi}{2}\}$, for fixed value of
$\theta_{3}=\frac{\pi}{6}$. Here, we assume $\alpha=2$. }
\end{figure}

In this section, we have plotted the success probability $P$ calculated
for both the ECPs in Sec. \ref{sec:ECP-for-partially} and Sec. \ref{sec:ECP-for-partially-1},
respectively. In Fig. \ref{fig:(Color-online)-(a)} (a), we have plotted
the variation of success probability $P$ with $\beta$ for our first
ECP, and it is shown that success probability $P$ can be controlled
by controlling coherent state parameters $\alpha$ and $\beta$. Specifically,
in Fig. \ref{fig:(Color-online)-(a)} (a), variation of $P$ with
$\beta$ is illustrated for $\alpha=0.5$, $1$, and $2$, to reveal
that the peak (maximum possible value) of the success probability
$P$ increases with the increase in $\alpha$ with a slight increase
in the corresponding value of $\beta$. Further, in our second ECP,
as $\beta^{2}+\gamma^{2}+\delta^{2}+\eta^{2}=1$ and $\beta,\gamma,\delta$
and $\eta$ are real, we can parameterize these parameters (i.e.,
$\beta,\gamma,\delta$ and $\eta$) in terms of new variables $\theta_{1}$,
$\theta_{2}$ and $\theta_{3}$ as $\beta=\cos[\theta_{3}]$, $\gamma=\sin[\theta_{3}]\cos[\theta_{2}]$,
$\delta=\sin[\theta_{3}]\sin[\theta_{2}]\cos[\theta_{1}]$, $\eta=\sin[\theta_{3}]\sin[\theta_{2}]\sin[\theta_{1}]$.
Subsequently, we plotted the variation of $P$ with $\theta_{1}$
and $\theta_{2}$ both varying from $0\rightarrow\frac{\pi}{2}$ for
$\theta_{3}=\frac{\pi}{6}$ and $\alpha=2$ as shown in Fig. \ref{fig:(Color-online)-(a)}
(b).

\section{Conclusion\label{sec:Conclusion}}

Entanglement is used in several quantum information processing applications.
In Sec. \ref{sec:Introduction}, we have already discussed the importance
of MES and how it is an ultimate resource for quantum information
applications. Unfortunately, it is a fact that these MES interact
with the environment over the time while processing, which leads to
the degradation in entanglement that lowers the efficiency of the
quantum communication schemes. Our goal is to avoid such degradation
in entanglement, and achieve the quantum communication applications
with high fidelity. To do so, many ECPs have been proposed and all
those existing ECPs have been designed for many different quantum
states \cite{ECP_Bennet,ECP_Bose,ECP_Shi,ECP_Zhao,ECP_Sheng1,ECP_Sheng2,ECP_Sheng3,ECP_Deng,Dhara_Cluster_ECP1,W-ECS,Our_ECP1,Our_ECP2,ECP_Wang,ECP_Sheng_2017,ECP_Sheng_2019}
as well as for Cluster state in different forms \cite{Lan_Cluster_ECP,Zhao_Cluster_ECP,Cluster_ECP2,Cluster_ECP5,Cluster_ECP6,Cluster_ECP7,Cluster_ECP8,Long_ECP_cluster,Song_Cluster_2017}.
However, to the best of our knowledge, no ECP has been reported for
4-mode Cluster-type ECS. This is the motivation for us to propose
ECPs for 4-mode Cluster-type ECS and it is the first ever ECP for
such state proposed by us. Another inspiring point is that, a Cluster-type
ECS \cite{Experimental_CV_Cluster1,Experimental_CV_Cluster} has extremely
important and interesting applications in the recent past \cite{QC_CV_Cluster1,QC_CV_Cluster2}.
Keeping this in mind, we have proposed two ECPs for 4-mode Cluster-type
ECS. A superposition of single-mode coherent state is used in the
first ECP while a superposition of single-mode coherent state and
a superposition of two-mode coherent state are used together in the
second ECP. To achieve the coherent state, we have used the parity
check measurement method, using a $50:50$ BS, i.e., balanced beam
splitter. We have also calculated the success probability $P$ of
each ECP in Sec. \ref{sec:Success-probability} and shown its variation
with the corresponding parameters in the Fig. \ref{fig:(Color-online)-(a)}
(a) and (b), respectively. Our ECPs have obtained the increased amplitude
$\sqrt{2}$ times higher in Eq. \ref{eq:Postselcetion} and \ref{eq: after_swap},
which could be advantageous for long distance quantum communication.
We conclude the paper with the anticipation that our ECPs would be
of practical interest and experimentally realizable with the present
linear optical technology. 

\textbf{Acknowledgments: }CS thanks to Tsinghua University, Beijing,
China for the postdoctoral fellowship support and the National Natural
Science Foundation of China under Grant No. 61727801. This work is
supported in part by the Beijing Advanced Innovation Center for Future
Chip (ICFC). CS also thanks to Anirban Pathak for the fruitful discussion
during her visit to JIIT, Noida, India.

\end{document}